\documentclass[%
 aip,
 amsmath,amssymb,
 reprint,%
]{revtex4-1}

\usepackage{graphicx}
\usepackage{dcolumn}
\usepackage{bm}

\usepackage[utf8]{inputenc}
\usepackage[T1]{fontenc}
\usepackage{mathptmx}
\usepackage{etoolbox}
\usepackage{color}
\usepackage{xcolor}

\makeatletter
\def\@email#1#2{%
 \endgroup
 \patchcmd{\titleblock@produce}
  {\frontmatter@RRAPformat}
  {\frontmatter@RRAPformat{\produce@RRAP{*#1\href{mailto:#2}{#2}}}\frontmatter@RRAPformat}
  {}{}
}%
\makeatother
\begin{document}

\preprint{AIP/123-QED}

\title{Resonance density range of absolute two-plasmon decay instability
}
\author{C. Yao}
\affiliation{ 
Department of Plasma Physics and Fusion Engineering and CAS Key Laboratory of Frontier Physics in Controlled Nuclear Fusion, University of Science and Technology of China, Hefei, Anhui 230026, People’s Republic of China
}%
\affiliation{Institute of Applied Physics and Computational Mathematics, Beijing 100088, People’s Republic of China}%

\author{J. Li}%
\thanks{Corresponding author: junlisu@ustc.edu.cn}
 \affiliation{ 
Department of Plasma Physics and Fusion Engineering and CAS Key Laboratory of Frontier Physics in Controlled Nuclear Fusion, University of Science and Technology of China, Hefei, Anhui 230026, People’s Republic of China
}%
\affiliation{ 
Collaborative Innovation Center of IFSA (CICIFSA), Shanghai Jiao Tong University, Shanghai 200240, People’s Republic of China
}%

\author{L. Hao}
\thanks{Corresponding author: hao\_liang@iapcm.ac.cn}
\affiliation{Institute of Applied Physics and Computational Mathematics, Beijing 100088, People’s Republic of China}%

\author{R. Yan}%
\affiliation{ 
Department of Modern Mechanics, University of Science and Technology of China, Hefei, Anhui
230026, People’s Republic of China
}%
\affiliation{ 
Collaborative Innovation Center of IFSA (CICIFSA), Shanghai Jiao Tong University, Shanghai 200240, People’s Republic of China
}%

\author{Q.Jia}%
 \affiliation{ 
Department of Plasma Physics and Fusion Engineering and CAS Key Laboratory of Frontier Physics in Controlled Nuclear Fusion, University of Science and Technology of China, Hefei, Anhui 230026, People’s Republic of China
}%

\author{Y-K. Ding}
\affiliation{%
Institute of Applied Physics and Computational Mathematics, Beijing 100088, People’s Republic of
China
}%

\author{J. Zheng}%
 \affiliation{ 
Department of Plasma Physics and Fusion Engineering and CAS Key Laboratory of Frontier Physics in Controlled Nuclear Fusion, University of Science and Technology of China, Hefei, Anhui 230026, People’s Republic of China
}%
\affiliation{ 
Collaborative Innovation Center of IFSA (CICIFSA), Shanghai Jiao Tong University, Shanghai 200240, People’s Republic of China
}%

\date{\today}

\begin{abstract}

We present a new insight into absolute two-plasmon decay (TPD) instability in nonuniform plasmas by identifying the resonance density range as the key parameter governing the growth of the resonant absolute modes. 
This range is defined as the density interval within which these resonant modes still exhibit growth in homogeneous plasmas. 
This range properly characterizes the spatial growth region of the resonant absolute modes in a series of linear fluid simulations across broad parameter spaces. 
Building on this insight, 
we investigate the absolute growth of TPD modes driven by laser pulses with intensity modulations, a common feature in broadband lasers used to suppress laser plasma instabilities. We establish the relationship between the resonance density range and the threshold time interval between intensity peaks, beyond which absolute growth is suppressed.


\end{abstract}

\maketitle

\section{introduction}

In pursuit of high-gain inertial confinement fusion (ICF), laser-plasma instabilities (LPIs) have been a critical issue that is necessary to be addressed~\cite{rosen2024,hurricane2023,craxton2015}. Among various LPI processes, two-plasmon decay (TPD) instability is one of the most prominent in direct-drive ICF~\cite{Seka2009a,Regan2010,Lian2025,LianCW2022}. TPD generates hot electrons that degrade the implosion by preheating the inner cold fuel before compression~\cite{Kirkwood2013a,baton2020,christopherson2021,christopherson2022}. Moreover, it leads to anomalous absorption of the pump laser, resulting in significant energy losses~\cite{turnbull2020}. Therefore, TPD need to be effectively suppressed to achieve high-gain ICF\cite{Sui2024,Zhang2023}.

Significant experimental progress~\cite{turnbull2020SSD,michel2013,A.Lei2024PRL} has been made in mitigating TPD and hot-electron generation.
In experiments employing traditional laser smoothing techniques, such as smoothing by spectral dispersion (SSD)~\cite{turnbull2020SSD} and polarization smoothing (PS)~\cite{michel2013}, partial suppression of TPD has been observed. 
Although reductions in hot- electron energy have been achieved under specific experimental configurations\cite{Follett2016}, the development of more generally
mitigation approaches is desirable to provide greater flexibility in experimental design.

Broadband lasers have been regarded as one of the most promising approaches for achieving this objective ~\cite{JJThomson1975,zhao2019,jiaxiaobao2023,Follett2019,MaHICMRE2021,wangxiangbing2024}. However, recent experimental results from the Kunwu broadband laser facility~\cite{JiLailin2020} reveal LPIs generate a higher yield of hot electrons compared to narrowband lasers ~\cite{Wang2024,A.Lei2024PRL}. 
The enhancement of three-halves harmonic emission and the observation of hot electron temperatures exceeding 40 keV support that TPD should be the primary mechanism responsible for the enhanced hot-electron generation~\cite{JunLi2025icmre}.
This unexpected enhancement has been attributed to intensity modulations intrinsic to broadband lasers~\cite{Yao2024,Liu2022,liuicmre2024}. The presence of high spikes in broadband laser fields limits the applicability of conventional TPD models \cite{Rosenbluth1973,Simon1983,afeyan1997a}, which are predominantly based on the plane-wave assumption.

Previous studies\cite{ZhangJ2014,yan2010} have shown that the growth of TPD is often initiated by absolute modes near the quarter-critical density ($n_c/4$). Owing to their low thresholds, these absolute modes can grow rapidly, propagate toward lower-density regions, and subsequently seed the amplification of convective modes \cite{ZhangJ2014}. 
Existing approaches for analyzing TPD absolute modes  involve complex mathematical derivations\cite{Simon1983,afeyan1997a,liu1976}, even under the idealized assumption of a plane-wave laser field. 
As a result, these methods exhibit clear limitations in addressing scenarios involving intensity modulation. 
Therefore, it would be valuable to enhance our physical understanding of absolute TPD growth and to develop new and simplified approaches.

In this study, we present new physical insights into the absolute TPD instability by demonstrating that the resonance density range of the absolute TPD mode governs its spatial growth. 
In a related work\cite{yao2025arXiv}, we explore the connection between the resonance density range and the nonlinear saturation levels of TPD. In the present article, however, we focus on its role in the linear growth of absolute TPD and its application to scenarios involving intensity-modulated laser pulses—a feature commonly associated with broadband laser systems.
A series of fluid simulations were conducted over a wide parameter space under conditions relevant to ICF. The simulation results show strong agreement between the resonance density range and the spatial growth region observed in the simulations.
We further examine absolute TPD driven by broadband laser pulses with intensity modulated at a frequency $\Delta \omega_m$. Under such conditions, the TPD modes exhibit intermittent growth synchronized with the modulation of the laser intensity, resulting in a modified instability threshold characterized by $\Delta \omega_g$. Building on this understanding, we provide a successful prediction of $\Delta \omega_g$.


The rest of the paper is organized as follows. In Sec.~II, the resonance density range is derived from linear theory, and it agrees with the growth regions observed in fluid simulations. In Sec.~III, fluid simulations of absolute TPD driven by laser pulses with intensity modulation are presented. The resonance density range is then applied to explain the modified instability threshold.
In Sec.~IV, we discuss the fundamental physical differences between convective and absolute instabilities. The main results are summarized in Sec.~V.

\section{Resonance density range of TPD modes}
\subsection{\label{sec:21} The resonance density range is derived from linear theory}


For a given electrostatic mode characterized by complex frequency $\omega = \omega_{ek} + i\gamma_0$ and wave vector $\vec{k}$, the resonance density range $n_r$ corresponds to the interval of electron density over which this mode act as an unstable TPD daughter wave. This range can be derived from the TPD dispersion relation for a homogeneous plasma \cite{Kruer2003}.

Here, we present the derivation of the resonance density range from the linear theory. The pump wave is assumed to be a monochromatic plane wave incident on the plasma, with its electric field polarized along the $y$ direction. The electric field of the incident laser is expressed as $\vec{E}_0 = E_0cos(\vec{k}_0\cdot \vec{x}-\omega_0t)\vec{e}_y$, where $E_0$ is the electricfield amplitude, $\vec{k}_0$ is the local laser wave vector, $\omega_0$ is the laser frequency, and $\vec{e}_y$ is the unit vectors along y direction. Based on the linearized TPD equations, the dispersion relation for TPD is obtained as \cite{Kruer2003}:

\begin{eqnarray}
&\left(\omega^2-\omega_{e k}^2\right)\left[\left(\omega-\omega_0\right)^2-\omega_{e,k-k_0}^2\right]= 
\nonumber\\
&\left[\frac{\mathbf{k} \cdot \mathbf{v}_{\mathrm{os}} \omega_{pe}}{2 k\left|\mathbf{k}-\mathbf{k}_0\right|}((\omega-\omega_0)k^2+\omega(\mathbf{k}-\mathbf{k}_0)^2)\right]^2
\label{eq:TPD_disper}   
\end{eqnarray} 
where the $\omega_{ek} = \sqrt{\omega_{pe}^2 + 3v_{te}^2k^2}$ represents the real part of $\omega$,  $\omega_{pe} = \sqrt{4\pi n_{e} e^2 / m_e}$ is the local plasma frequency, $n_e$ denotes the local electron density, $e$ and $m_e$ are the electron charge and mass, respectively. The $v_{te} = \sqrt{T_e / m_e}$ represents the electron thermal velocity, with 
$T_e$ being the electron temperature. Additionally, $v_{os}$ is 
the electron oscillation velocity in the laser wave field. 

\begin{figure}
\includegraphics[width=0.95\linewidth]{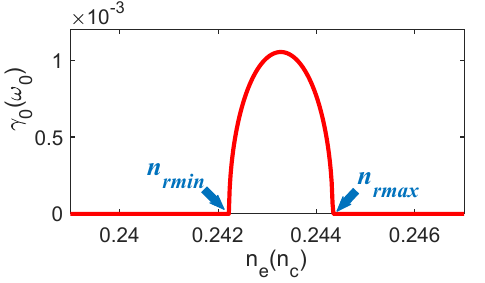}
\caption{\label{fig:nr} 
Linear growth rate $\gamma_0$ of the resonant mode with $(k_{xr} \approx 0.88\,\omega_0/c,\ k_{yr} \approx 0.06\,\omega_0/c)$ as a function of plasma density. Parameters: $L_n = 300 \mu m$, $I = 3.0 \times 10^{14}~\text{W}/\text{cm}^2$,$\lambda_0 = 1/3\mu m$, and $T_e = 3.0~\text{keV}$.
}
\end{figure}

By scanning the electron density $n_e$, the linear growth rate $\gamma_0$ of a mode is obtained as a function of electron density $n_e$ using Eq.~(\ref{eq:TPD_disper}). Figure~\ref{fig:nr} shows the variation of the growth rate $\gamma_0$ for the mode $\vec{k}_r = (k_{rx}, k_{ry}) \approx(0.88\omega_0/c, 0.06\omega_0/c)$ as a function of $n_e$, where c is the speed of light in vacuum. This resonant mode, derived from Simon's theory, is the absolute mode with the maximum growth rate under the parameters $L_n = 300 \mu m$, $I = 3.0 \times 10^{14}W/cm^2$, $\lambda_0 = 1/3\mu m$, and $T_e = 3.0 keV$. The results reveal that this mode exhibits a positive growth rate $\gamma_0$ within the density interval $n_r = (n_{r\min}, n_{r\max})=(0.2422n_c,0.2443n_c)$, which defines the resonance density range. This range corresponds to a resonance width $\Delta n_r \approx 0.0021n_c $. The underlying physical interpretation is straightforward: the mode can grow only in regions where it remains linearly unstable.

\subsection{\label{sec:22} Growth regions of absolute TPD from fluid simulations}

We perform a series of two-dimensional (2D) fluid simulations using the LTS code\cite{yan2010}, wihch solves the linear TPD equations. The LTS code has been successfully benchmarked with established TPD theories\cite{Simon1983,afeyan1997a} for cases with linear density profiles\cite{yan2010}. The physical parameters employed in the simulations are typical OMEGA experiments, as summarized in Table.\ref{tab:simu_single}. The electron density profile $n_e(x)$ of the CH plasma is linear along the x-direction, spanning the range $0.18n_c\sim 0.28n_c$, where $n_c$ is the critical density for the laser pulses with wavelength $\lambda_0 = 1/3 ~\mu m$. 
The laser pulses are launched as external electric fields in the form of linearly polarized plane waves (polarized along the $y$-direction), and propagate in the $x$-direction. Absorbing boundary conditions for the electrostatic plasma waves are applied along the $x$-direction, while periodic boundary conditions are used along the $y$-direction.

\begin{table}
\caption{\protect\label{tab:simu_single} Initial simulation parameters—density scale length ($L_n$), electron temperature ($T_e$), laser intensity ($I_0$), and the TPD threshold factor $\eta = I L_n \lambda_0 / 81.86T_e$—with $\lambda_0 = 1/3\mu m$ for all cases.}
\begin{ruledtabular}
\begin{tabular}{ccccccccc}
 Index&$L_n$ &$T_e$  &$I_0$ & $\eta$ \\
 &($\mu$m)&(keV)&($\times 10^{14}W/cm^2$) \\
\hline
(i)& 300 & 3.0  &3.0& 1.2 \\
(ii)& 300 & 3.0  &6.0& 2.4 \\
(iii)& 150 & 3.0  &6.0 & 1.2 \\
(iv)& 200& 3.0 &4.5& 1.2 \\ 
(v)& 150 & 2.5  &6.0& 1.5 \\
(vi)& 150 & 1.0  &6.0 & 3.7 \\
\end{tabular}
\end{ruledtabular}
\end{table}

\begin{figure}
\includegraphics[width=1.0\linewidth]{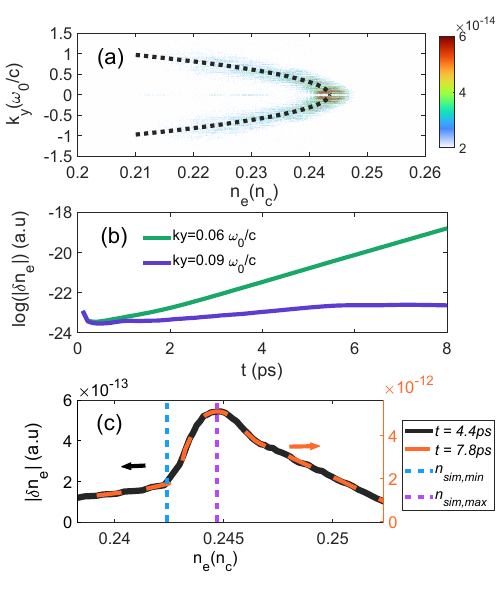}
\caption{\label{fig:mode_range}Simulation results of case(i) in Table.\ref{tab:simu_single}. (a) Typical $k_y - n_e$ spectra of $\delta n_e$ (at $t=1.2 ps$), the black dotted line represents the characteristic TPD parabola. (b) Time-evolving behavior of two different modes : $k_y = 0.06 ~\omega_0/c$ is absolute TPD mode, $k_y = 0.09~\omega_0/c$ is convective mode. (c) the spatial envolope of $k_y = 0.06 ~\omega_0/c$ at $4.4 ps$ (black solid line) and $7.8 ps$ (orange dotted line). The corresponding density of vertical purple dot line is density $n_e = 0.2424n_c$, and the green dot line is correspond to $n_e = 0.2447n_c$.}
\end{figure}

By performing a Fourier transform along the $y$-direction on the spatial distribution of the electron density perturbation $\delta n_e$, we obtain the $k_y$-$n_e$ spectrum of $\delta n_e$ from the simulations. A representative spectrum is presented in Figure~\ref{fig:mode_range}(a). All observed modes lie on the characteristic parabola associated with TPD~\cite{Kruer2003}[the black dotted line in Figure~\ref{fig:mode_range}(a)]. The modes with larger $k_y$ appear in the lower-density region, corresponding to convectively unstable modes, whereas those with smaller $k_y$ reside in the higher-density region and are associated with absolutely unstable modes. These behaviors are consistent with those reported in previous studies~\cite{yan2012,yan2009}.

We focus on investigating the temporal and spatial evolution of the amplitude $\delta n_e$ of the absolute resonant mode. By averaging the spectral signal corresponding to a given $k_y$ over the $x$-direction from the $k_y$-$n_e$ spectrum, we obtain the temporal evolution of the amplitude for each unstable mode. To extract the spatial evolution, we filter the spectral component at a selected $k_y$ from the $k_y$-$n_e$ spectrum, thereby obtaining the spatial distribution of the mode amplitude $\delta n_e$ along the $x$-axis.

The results in Figure 2(b) reveal two representative types of temporal evolution for unstable modes. For the convective mode with $k_y \approx 0.09 \omega_0/c$, represented by the purple curve, the amplitude initially grows and then saturates. In contrast, the absolute modes emerge closer to $0.25 n_c$, and therefore exhibit smaller $k_y$ compared to the convective modes. As shown by the green line, the absolute mode with $k_y \approx 0.06 \omega_0/c$ exhibits exponential growth over time.


As the absolute mode undergoes continuous exponential growth in the absence of any saturation mechanisms in the simulation, its spatial envelope develops into a single-peaked structure located near the density predicted by TPD theory.
Figure~\ref{fig:mode_range}(c) presents the $\delta n_e$ envelopes of the resonant mode ($k_y \approx 0.06\omega_0/c$) under the parameter settings of case~(i) in Table~\ref{tab:simu_single}, at two different simulation times, plotted with different scales. The two envelope curves almost overlap, indicating that the mode shape (not amplitude) has reached a steady state.

The presence of the peak reveals that the instability is localized, with both daughter waves propagating and simultaneously amplifying each other within the amplification region.
The forward-propagating daughter wave possesses a relatively large group velocity along x-direction, whereas the backward-propagating wave has a nearly zero $k_x$, resulting in its propagation primarily along the $y$-direction. Consequently, the spatial $\delta n_e$ envelope [Figure.~\ref{fig:mode_range}(c)] describes the amplitude growth of the forward-propagating daughter wave during its propagation. The spatial region over which this growth occurs represents the growth range of the absolute mode.


Based on this analysis, the growth region of an absolute mode in simulations is defined as follows: (i) the peak of the $\delta n_e$ envelope marks the higer density boundary ; (ii) the lower density boundary is defined by the point at the valley of the $\delta n_e$ profile where the second derivative attains its maximum. For the unstable mode with $k_y = 0.06\omega_0/c$ in case~(i), the density growth range extracted from simulation is $n_{\mathrm{sim}} = (0.2424n_c, 0.2447n_c)$, yielding a growth range of $\Delta n_{\mathrm{sim}} = 0.0023n_c$, agreeing well with the theoretically predicted resonance range density $\Delta n_r = 0.0021n_c$ for this case.

\subsection{\label{sec:23} Resonance density range characterizes the growth range of absolute TPD mode}

\begin{figure}
\includegraphics[width=0.95\linewidth]{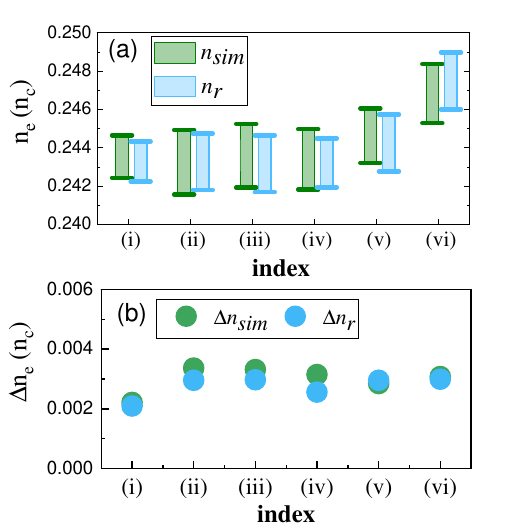}
\caption{\label{fig:diff_LTS_simulation}(a)Compare rensonance density ranges $n_r$ with growth ranges from simulations of Table. \ref{tab:simu_single}.(b)Compare $\Delta n_r$ with $\Delta n_{sim}$.}
\end{figure}

For all the simulations in Table \ref{tab:simu_single}, we compare the resonance density range $n_r$ with the simulated growth region $n_{\mathrm{sim}}$, as illustrated in Figure.~\ref{fig:diff_LTS_simulation}(a). The blue bars represent the theoretically predicted resonance density ranges $n_r$, while the green bars correspond to $n_{\mathrm{sim}}$ extracted from simulations. Across a wide range of parameters, $n_{\mathrm{sim}}$ shows good agreements with the calculated $n_r$, and their widths, $\Delta n_{\mathrm{sim}}$ and $\Delta n_r$, are nearly identical, as shown in Figure.~\ref{fig:diff_LTS_simulation}(b).

These results demonstrate that the resonance density range $n_r$ properly characterizes the spatial growth region of the resonant absolute modes, and further support that absolute modes growth are confined to regions where the modes remain linearly unstable.

The preceding theoretical analysis was conducted under idealized conditions without considering damping effects. However,
in realistic scenarios, damping effects must be considered, which tend to reduce the effective growth region. When collisional damping is taken into account, the actual growth region corresponds to the density interval where the linear growth rate $\gamma_0$ exceeds the electron-ion collision frequency $\nu_{ei}$. Landau damping, on the other hand, remains negligible due to small $k\lambda_D$ ($\lambda_D$ is the Debye length) within the relevant density range for absolute instabilities and is typically neglected in such analyses. 
Nevertheless, since the growth rate $\gamma_0$ varies steeply with $n_e$ [Figure~\ref{fig:nr}], replacing the condition $\gamma_0 > 0$ with $\gamma_0 > \nu_{ei}$ results in only a small variation in $n_e$. As a result, the damping effects on $n_r$ are minimal and does not significantly alter the conclusions drawn from linear theory.

\section{Applications of resonance density range: Absolute TPD instability driven by intensity modulated laser pulses}

Broadband lasers have been considered one of the most promising approaches to mitigate  LPIs~\cite{thomson1974,JJThomson1975,zhao2019,Follett2019}. Experiments conducted at the Kunwu facility~\cite{A.Lei2024PRL,Wang2024} ($\Delta\omega/\omega_0\sim 0.57\%$)  have demonstrated that broadband lasers can effectively suppress stimulated Brillouin scattering (SBS), and also exhibit a certain degree of suppression of stimulated Raman scattering (SRS) depending on the laser plasma parameters and configurations. However, the enhanced  hot electrons and emission of the $3\omega_0/2$ scattered light under broadband laser suggests that bandwidth may, in fact, enhance TPD.

Previous studies~\cite{Yao2024} suggest that the enhanced generation of hot electrons can be caused by the significant intensity modulations of broadband laser fields\cite{Yao2024}. In such modulated filed, TPD grows intermittently and leads to a burst of hot electrons near the intensity peaks, resulting in a higher overall hot-electron energy fraction compared to narrowband laser.

To investigate the intermittent growth of TPD induced by intensity modulation, it is essential to accurately characterize the temporal waveforms of intensity-modulated broadband laser fields. However, obtaining such detailed characterizations remains a significant challenge. Given that these fields typically consist of numerous modulation structures with varying amplitudes and durations \cite{JiLailin2020}, a fundamental first step is to examine how the amplitude and duration of intensity modulations individually influence the instability growth.

In this section, a simplified two-color laser model is employed to conveniently control the modulation frequency $\Delta\omega_m$, enabling a systematic investigation of absolute TPD responses across a specified range of $\Delta\omega_m$.
A series of linear fluid simulatinos using LTS code is performed, scanning representative laser–plasma parameters relevant to ICF conditions.

\subsection{\label{sec:31}Simulation setup}

We employ a two-color laser model to construct a wave field that is periodically modulated at a frequency $\Delta\omega_m$. The field consists of two frequency components, $(\omega_0, k_0)$ and $(\omega_0', k_0')$, each with equal amplitude $E_0$. The frequency difference between these components satisfies $\Delta\omega_m = \omega_0 - \omega_0'$. This study primarily considers the regime $\Delta\omega_m \ll \omega_0$, in which case the intensity envelope $I_m$ oscillates at the modulation frequency $\Delta\omega_m$ and can be expressed as\cite{Yao2024}:






\begin{equation}
I_m(x, t,\Delta \omega_m) = I_0 \left[1 + \cos(\Delta k_m x - \Delta \omega_m t)\right]
\label{eqn:Im}
\end{equation}

where $I_0$ is the time-averaged value of $I_m(x,t,\Delta\omega_m)$ at any space location, and $\Delta k_m=k_0-k_0'$. And the random phase is neglected in the formulation, as it has no impact on the derivation or the resulting conclusions\cite{Yao2024}.

The simulation setup in this section is similar to that described in Section~\ref{sec:22}, except that the pump lasers are replaced with two-color drivers. The physical parameters used in these simulations are summarized in Table~\ref{tab:simu_twocolor}. In each series of simulations,  $\Delta\omega_m$ is scanned from  $0.1\%\omega_0$ to $3.0\%\omega_0$.


\begin{table}
\caption{\protect\label{tab:simu_twocolor} Initial simulation parameters ($L_n$, $T_e$, $I_0$) for two color drivers.}
\begin{ruledtabular}
\begin{tabular}{ccccccccc}
 Index&$L_n$ &$T_e$  &$I_0$  \\
 &($\mu$m)&(keV)&($\times 10^{14}W/cm^2$) \\
\hline
(A)& 100 & 3.0  &17 \\
(B)& 150 & 3.0  &11  \\
(C)& 200& 3.0 &9.0 \\ 
(D)& 250 & 3.0  &7.0 \\
(E)& 300 & 3.0  &6.0  \\
\end{tabular}
\end{ruledtabular}
\end{table}

\subsection{\label{sec:32} Results: growth threshold $\Delta\omega_g$ and decoupling threshold $\Delta \omega_s$ for absolute TPD}


In the cases of two-color drivers, the resonant absolute mode is defined as the unstable mode with the maximum $\gamma$ from simulation. Under the same plasma conditions, the resonant modes are found to be nearly identical between the single-color and two-color cases. The resonant modes are found to be nearly independent of $\Delta\omega_m$, remaining essentially unchanged across the entire range explored.

\begin{figure}
\includegraphics[width=0.96\linewidth]{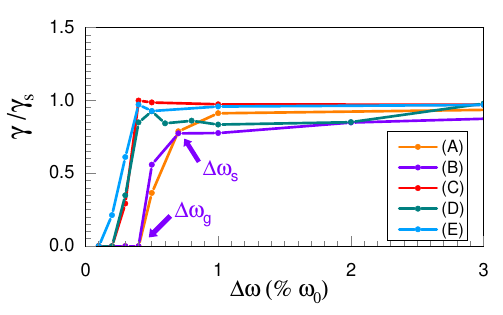}
\caption{\label{fig:all_two} Fluid Simulation results of Table \ref{tab:simu_twocolor}. The normalized growth rate $\gamma/\gamma_{s}$ of two-color driver varies with modulated frenquency $\Delta \omega_m$ for case(A) to case(E).The marked $\Delta\omega_g$ and $\Delta\omega_s$ correspond to case (B).}
\end{figure}

Following the method described in the previous section, we obtain the temporal growth rate $\gamma$ of the resonant mode for a series of simulations, in which $\Delta\omega_m$ is scanned from $0.1\% \omega_0$ to $3.0\%\omega_0$ for cases~(A) through~(E) in Table~\ref{tab:simu_twocolor}.
 As shown in Figure~\ref{fig:all_two}, all values of $\gamma$ are normalized to $\gamma_{s}$,  where $\gamma_{s}$ is the absolute TPD growth rate driven by a single-frequency laser with intensity $I_0/2$, and $I_0$ represents the average intensity of the two-color drivers.

The simulation results for all the parameter sets exhibit a similar trend in Figure~\ref{fig:all_two}. In the limit of small $\Delta\omega_m$, the TPD instability is suppressed. As $\Delta\omega_m$ increases beyond the first threshold $\Delta \omega_g$, the absolute TPD modes begin to grow; however, the corresponding growth rate $\gamma$ remains below $\gamma_s$. Once $\Delta\omega_m$ exceeds the second threshold $\Delta \omega_s$, $\gamma$ approache $\gamma_s$. In this regime, the results are consistent with previous studies~\cite{zhao2017}, which indicate that the two frequency components become effectively decoupled and independently drive the TPD instability, resulting in a growth rate equivalent to that of a single-frequency driver with half the average incident intensity.



\subsection{\label{sec:33} Mechanism: thresholds of intermittently excited absolute TPD}


\begin{figure*}
 \includegraphics[width=0.70\linewidth]{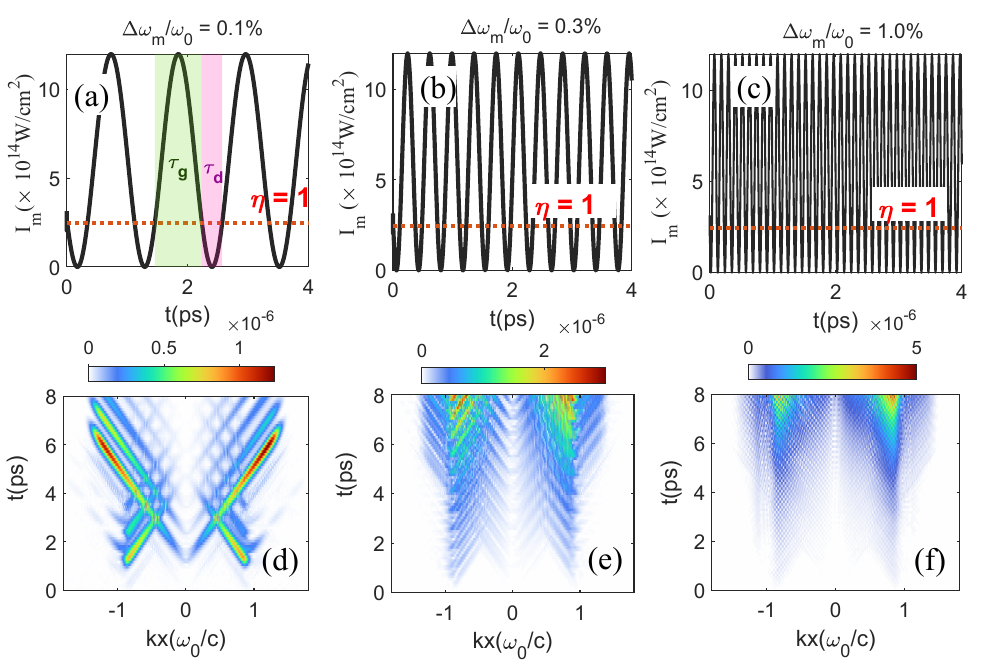}
 \caption{\label{fig:diff_wm} Simulation results of case(E) in Table~\ref{tab:simu_twocolor}.  $(a)(d)$correspond to $\Delta\omega_m/\omega_0=0.1\%$, $(b)(e)$correspond to $\Delta\omega_m/\omega_0=0.3\%$, $(c)(f)$correspond to $\Delta\omega_m/\omega_0=1.0\%$.  $(a)\sim(c)$ the incident intensity envelopes $I_m(t)$, with the red dashed lines indicating the TPD threshold $\eta = 1$. $(d)\sim(f)$ the temporal evolution of the $k_x$ spectra of  $|\delta n_e|$ at $k_y = 0.06\omega_0/c$.}
\end{figure*}

\begin{figure*}
 \includegraphics[width=0.75\linewidth]{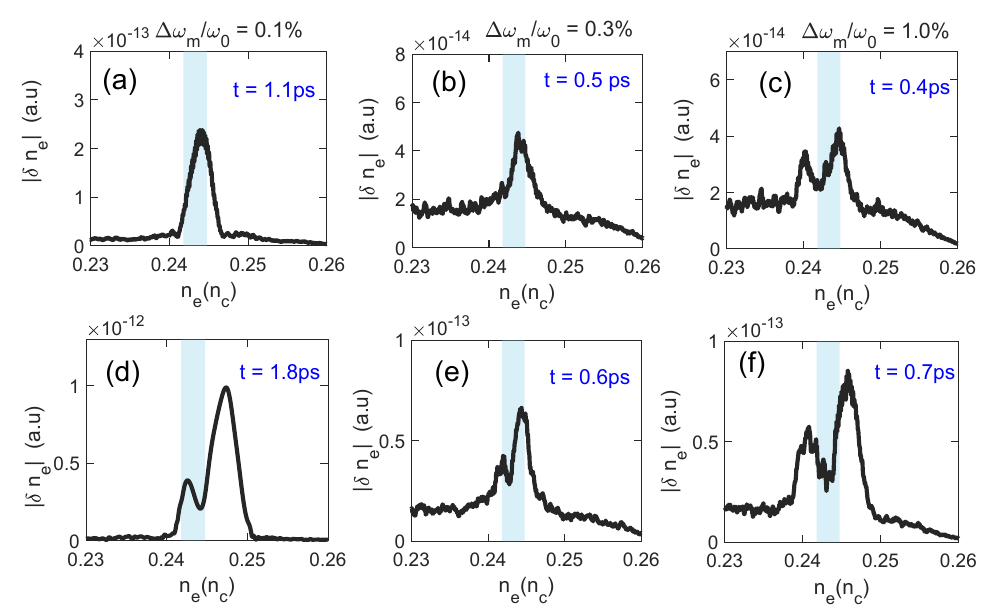}
 \caption{\label{fig:diff_wm_TPDmode} The $|\delta n_e|$ special envolopes at $k_y = 0.06~\omega_0/c$ from simulation results of case(E) in Table~\ref{tab:simu_twocolor}. The blue shaded regions represent the resonance density range $n_r=(0.2418n_c, 0.2447n_c)$. $(a)(d)$ correspond to $\Delta\omega_m/\omega_0=0.1\%$, $(b)(e)$correspond to $\Delta\omega_m/\omega_0=0.3\%$, $(c)(f)$correspond to $\Delta\omega_m/\omega_0=1.0\%$. }
\end{figure*}


To investigate the temporal evolution of the absolute TPD modes, we obtain the temporal $k_x$ spectra of the electron density perturbations $\delta n_e$ associated with the resonant absolute TPD modes, as shown in Figure~\ref{fig:diff_wm}$(d)\sim (f)$.   


Figure~\ref{fig:diff_wm} $(a)\sim(c)$ show the time-modulated laser intensity $I_m(t, \Delta\omega_m)$ for case~(E) in Table~\ref{tab:simu_twocolor}, corresponding to $\Delta\omega_m = 0.1\%\omega_0$ in (a), $\Delta\omega_m = 0.3\%\omega_0$ in (b) and $\Delta\omega_m = 1.0\%\omega_0$ in (c). In these laser schemes, the intensity is modulated with a period of $T = 2\pi/\Delta\omega_m$, resulting in a peak intensity $I_{peak}$ that is twice the average value $I_0$, and intensity valleys that approach zero. As the modulation frequency $\Delta\omega_m$ decreases, both the peak and valley durations last for longer time; conversely, higher $\Delta\omega_m$ results in shorter peak and valley durations.

The temporal evolution of the $k_x$ spectra at the resonant absolute modes [Figure~\ref{fig:diff_wm} $(d)\sim(f)$] reveals the varying effects of $\Delta \omega_m$ on the growth of TPD. For $\Delta\omega_m/\omega_0 = 0.1\%$, Figure~\ref{fig:diff_wm}(d) displays that the absolute TPD grows intermittently. The first peak of the absolute TPD modes intensity arises at approximately $1.1ps$, while the second peak occurs at around $2.2ps$, resulting in a phase interval of about $1.1ps$ between the two peaks, which is in good agreement with the modulation period of the laser, $T = 2\pi/\Delta\omega_m \approx 1.1 ps $.This implies that, in this case, the growth of absolute TPD is predominantly governed by the temporal modulation of the laser pulse. Notably, each modulation period contains both a growth phase and a decay phase of the mode, which represents a distinctive characteristic of the dynamics. 


This intermittent behavior arises from the fact that TPD can only be excited when the instantaneous laser intensity $I_m(t)$ exceeds a threshold $\eta$, where $\eta = IL_n\lambda_0 / 81.86T_e$, as indicated by the red dotted lines in Figure~\ref{fig:diff_wm}$(a)\sim (c)$. When $I_m(t)$ falls below $\eta$, the laser intensity is insufficient to sustain TPD growth. Meanwhile, the modes excited by the previous peaks continuously propagate out of the resonant region, leading to a decrease in its amplitude within the resonant density range.

This suggests that the effect of intensity modulation on absolute TPD can be separated into two distinct contributions: a growth phase $\tau_{g}$ [the green shaded region in Figure \ref{fig:diff_wm}(a)]  dominated by intensity peaks, yielding an effective growth rate $\gamma_{eff}$. And a decay phase $\tau_{d}$ [the red shaded region in Figure \ref{fig:diff_wm}(a)] near the intensity valleys, governed by an effective decay rate $\nu_{eff}$, which is associated with the propagation velocity of TPD. The interplay between the growth and decay phases of this mode determines whether it will ultimately become unstable.

Different $\Delta\omega_m$ leads to variation in $\tau_{g}$ and $\tau_{d}$. Specifically, a smaller $\Delta\omega_m$ results in a longer $\tau_d$, allowing the already amplified modes to propagate farther and thereby reducing their ability to provide stronger seeds for subsequent growth. Consequently, in the limit of small $\Delta\omega_m$, the absolute TPD mode ultimately fails to grow.

For instance, in the case of $\Delta\omega_m = 0.1\%\omega_0$ [Figure~\ref{fig:diff_wm}(d)], the TPD mode exhibits a growth phase of approximately $\tau_g \approx 0.5~ps$ and a longer decay phase of $\tau_d \approx 0.6 ps$. As shown in Figure~\ref{fig:diff_wm_TPDmode}(a), at $t = 1.1~ps$, the first peak of this mode is in its growth process. Within a single modulation period, $T = 2\pi/\Delta\omega_m \approx 1.1~\mathrm{ps}$, the second peak emerges near $t = 1.8~\mathrm{ps}$. By this time, the main part of the first peak has already propagated out of the resonant region [indicated by the blue shaded area in Figure~\ref{fig:diff_wm_TPDmode}(d)]. As a result, the preceding peak cannot seed the subsequent one, causing the growth of each successive peak effectively decoupled from that of its predecessor and preventing it from exhibiting continuous growth.

As $\Delta\omega_m$ gradually increases, the main part of the preceding peak remains within the resonance density range due to the short $\tau_d$, thereby seeding the growth of the subsequent one. As a result, successive peaks couple with their predecessors. Consequently, once $\Delta\omega_m$ reaches $\Delta\omega_g$, where $\Delta\omega_g$ represents the threshold at which the absolute TPD modes become unstable, the absolute TPD instability can further grow.
As shown in Figure~\ref{fig:diff_wm_TPDmode}(e), for $\Delta\omega_m = 0.3\%\omega_0$, the second peak rises while a significant portion of the first peak still remains within the resonance density range (indicated by the blue shaded region). Consequently, the first peak provides a strong seed for the second peak, allowing this mode to amplify to a higher amplitude than the first peak. Such a recurring process progressively enhances the mode’s amplitude, ultimately causing it to become unstable. In this case, the mode still retains the characteristic of intermittent growth [Figure~\ref{fig:diff_wm}(e)], with a modulation-period phase of $0.4~\mathrm{ps}$, which is close to the intensity modulation period $T = 2\pi/\Delta\omega_m \approx 0.4~\mathrm{ps}$.


As $\Delta\omega_m$ continues to increase until it exceeds the threshold $\Delta\omega_s$, where $\Delta\omega_s$ corresponds to the point at which $\gamma$ approaches $\gamma_s$, the modulation period becomes so short that the TPD modes can no longer respond to the modulation, causing the characteristic of intermittent growth to vanish [Figure~\ref{fig:diff_wm}(f)]. In this regime, Figures~\ref{fig:diff_wm_TPDmode}(c) and (f) show that the growth regions of the two frequency components become spatially separated, enabling them to drive TPD independently throughout the entire process. As a result, the TPD mode exhibits sustained exponential growth, with the growth rate $\gamma$ reaching $\gamma_{s}$ [Figure~\ref{fig:all_two}].

\subsection{\label{sec:34} Modeling the growth threshold $\Delta\omega_g$}

The evaluation of $\Delta \omega_g$ will improve future analyses of the effect of laser intensity modulation peaks in broadband lasers on TPD.
We now investigate the dependence of $\Delta\omega_g$ on the resonance density range $n_r$ and other physical parameters. As outlined in Section~\ref{sec:23}, $n_r$ defines the region where the mode is linearly unstable, implying that absolute growth is possible only within this range. 
Furthermore, as discussed in section \ref{sec:33}, in the limit of small $\Delta\omega_m$, the mode largely propagates out of the resonant region $n_r$ during the long decay phase, and cannot provide a strong seed for the subsequent growth. As $\Delta\omega_m$ increases, the mode seeded by the preceding intensity peak remains partially within the resonant region, allowing it to be further driven by the subsequent peak [Figure~\ref{fig:diff_wm_TPDmode}(e)]. This overlap allows the next peak to excite higher-amplitude EPWs on a pre-perturbed density background.
The corresponding frequency gap at this threshold is defined as $\Delta\omega_g$.

Based on this concise physical picture, the escape time $\tau_c$ ($\Delta\omega_g=2\pi/\tau_c$) of EPWs from the resonance density range $n_r$ can be calculated. This time depends on the group velocities ($v_{g1}$ and $v_{g2}$) of the two TPD daughter waves, as well as the density scale length $L_n$. For a given resonance width $\Delta n_r$, a larger group velocity leads to a shorter escape time $\tau_c$, while a longer $L_n$ corresponds to a longer $\tau_c$. 

The progation time can be estimated based on their separate group velocities as follows:

\begin{equation}
\label{eq:tp}
    \tau_{p1} \sim \frac{\Delta n_r L_n}{v_{g1}}\quad\quad
    \tau_{p2} \sim \frac{\Delta n_r L_n}{v_{g2}}
\end{equation}

We reasonably have the scaling relation $\tau_c \sim \tau_{p1} \cdot \tau_{p2}$, which in turn yields the critical modulation frequency $\Delta\omega_g \sim v_{g1}v_{g2}/(\Delta n_r L_n)^2$.

We now consider the influence of the laser intensity $I_0$, as the instability exhibits strong sensitivity to this parameter during its dynamic growth. The effect of laser intensity on the system can be categorized into two main aspects.

First, higher laser intensity increases the effective growth rate $\gamma_{eff}$, enabling unstable modes to reach higher amplitudes during $\tau_g$ phase. 
As shown in Figure~\ref{fig:diff_wm_TPDmode}(a), laser intensity modulation leads to a spatial envelope of the TPD mode composed of approximately Gaussian-shaped peaks.  As the peak amplitude increases, the trailing part of the mode also rises. This results in the trailing edge, which has not fully propagated out of the resonance density region, provide stronger seed amplitudes for the next TPD growth when the following intensity peak arrives, enabling it to reach a higher amplitude. Consequently, a higher $I_0$ promotes the growth of TPD modes and reduces the threshold modulation frequency, following the scaling $\Delta\omega_g \sim 1/\gamma$.

Second, increasing $I_0$ extends the effective growth phase $\tau_g$, related to the duration during which the laser intensity exceeds a threshold level $\eta$. Thus, $\eta$ serves as an effective indicator of the growth window. A larger $\eta$ corresponds to a longer $\tau_g$, which allows EPWs to grow to higher amplitudes and shortens their decay phase $\tau_d$. As a result, the growth threshold frequency decreases accordingly, with the scaling relation $\Delta\omega_g \sim 1/\eta$.

 Based on the above analysis, we propose the dependence of $\Delta\omega_g$ as follows:

\begin{equation}
\label{eq:omega_c}
    \Delta\omega_g \sim  \frac{v_{g1}v_{g2}}{(\Delta n_r L_n)^2}\cdot \frac{1}{\gamma}\cdot \frac{1}{\eta} \cdot \alpha
\end{equation}

where $\alpha$ serves as an adjustable parameter. We adopt the Simon growth rate, evaluated at the average laser intensity $I_{0}$, as an approximation for the growth rate $\gamma$ of the resonant mode. 

\begin{figure}
\includegraphics[width=0.90\linewidth]{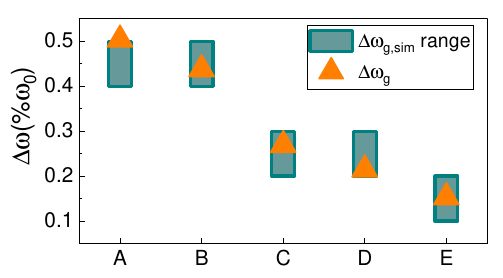}
\caption{\label{fig:diff_Dwc} Comparison of $\Delta\omega_g$ from Eq.~(\ref{eq:omega_c}) and $\Delta\omega_{g,sim}$ from fluid simulations. }
\end{figure}

By applying the scaling relation in Eq.(\ref{eq:omega_c}), $\Delta\omega_g$ can be evaluated for each simulation case. With $\alpha \approx 50$, Eq.(\ref{eq:omega_c}) provides reasonable estimates of $\Delta\omega_g$ across a wide range of laser–plasma conditions. As illustrated in Figure~\ref{fig:diff_Dwc}, the estimated values are in good agreements with the simulation results $\Delta\omega_{g,\text{sim}}$.



\section{Discussions: Effects of resonance density range on convective TPD}

According to Eq.~(\ref{eq:TPD_disper}),
the resonance density range is valid for all TPD modes, including both absolute and convective modes. In previous discussions, we focus on the effects on absolute TPD modes near the  $n_c/4$. In this section, we turn to convective TPD modes, which occur off the $n_c/4$ density, mainly between $0.21n_c$ and $0.235n_c$. For convective instability, the unstable modes grow during propagation along the density gradient. The total gain is defined as the amplification factor within a growth range where the phase mismatch caused by inhomogeneous density remains tolerable. Therefore, the comparison between the convective growth range and resonance density range is important to analyze the role of resonance density range in convective instability.

\begin{figure}
\includegraphics[width=0.90\linewidth]{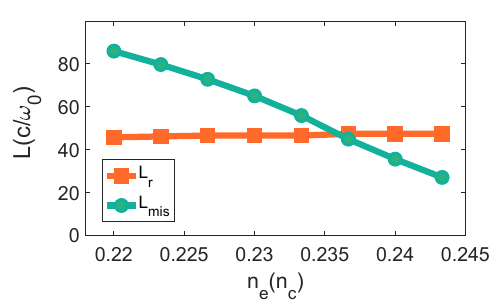}
\caption{\label{fig:diff_L} Mismatch length $L_{\rm mis}$ and resonance length $L_r$ for case (i) in Table~\ref{tab:simu_single} as functions of the electron density $n_e$. }
\end{figure}

 Figure~\ref{fig:diff_L} illustrates two characteristic length scales associated with the three-wave interaction process.  The resonance length $L_r$ is derived from the resonance density range $\Delta n_r$ and is given by 
$L_r = \Delta n_r /(d n_e/dx)= \Delta n_rL_n/(n_c/4)$. The mismatch length $L_{\mathrm{mis}}$ is defined as the length over which the phase mismatch $\int \kappa dx$ accumulates to $\pi/2$, where $\kappa$ is the spatial derivative of the phase mismatch $\Delta k$. Here, $\Delta k = k_0 - k_{x1} + k_{x2}$ and $\kappa = d\Delta k/dx$. The $k_{x1}$ and $k_{x2}$ represent the $x$-components of the wave vectors of the two TPD daughter waves. %


Figure.~\ref{fig:diff_L} shows that different characteristic lengths dominate the growth region for convective and absolute TPD. The resonance length $L_r$ varies weakly with plasma density, whereas $L_{mis}$ decrease with increasing density. 
Notably, in the region where the electron density $n_e$ is below $0.237n_c$, $L_{mis}$ exceeds $L_r$, suggesting that convective amplification dominates the overall gain in this range. Conversely, for $n_e > 0.237n_c$, $L_r$ becomes larger than $L_{mis}$, and the overall growth behavior is primarily governed by the resonance density range.

These observations suggest that distinct mechanisms govern the growth of convective and absolute TPD. The growth of convective TPD is limited by phase mismatch, whereas the growth of absolute TPD is determined by the resonance density range. \textcolor{black}{As this density range for absolute TPD ($\sim0.003n_c$ from Fig.~\ref{fig:diff_LTS_simulation}) is much shorter compared to the entire density range of TPD growth, the TPD spectra at saturation are usually dominated by convective TPD modes\cite{yan2009,ZhangJ2014}.}

\section{Summary}

\textcolor{black}{In summary, }we propose a new physical sight of absolute TPD instability by identifying the resonance density range $n_r$ as the key parameter governing its spatial growth. This range effectively characterizes the spatial growth region of absolute modes, as demonstrated by a series of linear fluid simulations across a broad parameter space. This insight allows us to explain the growth threshold of absolute TPD under laser pulses with intensity modulation, and to develop a predictive model for the dependence of the growth threshold $\Delta\omega_g$ on the newly defined resonance density range $n_r$.
\textcolor{black}{Future work will investigate the effects of oblique incidence and polarization on absolute TPD growth from broadband lasers to better align with experimental conditions. }



\begin{acknowledgments}
This research was supported by the National Natural Science Foundation of China (NSFC) (Grant Nos. 12275269, 12388101, 12375243, 12275032 and U2430207), by the Strategic Priority Research Program of Chinese Academy of Sciences (Grant Nos. XDA25010200 and XDA25050400). The numerical simulations reported in this paper were performed at the Hefei Advanced Computing Center.
\end{acknowledgments}

\appendix

\section*{References}
\bibliographystyle{iopart-num}
\providecommand{\noopsort}[1]{}\providecommand{\singleletter}[1]{#1}%

\end{document}